\documentclass{jpp}
\usepackage{graphicx}
\usepackage{epstopdf, epsfig}
\usepackage{hyperref}

\usepackage{amsfonts}
\usepackage{times}
\usepackage{graphicx}
\usepackage{natbib}
\usepackage{color}

\renewcommand{\vec}[1]{\mbox{\boldmath $#1$}}

\def \Om  {{\it \Omega}}

\def \Pm {\ensuremath{\rm{Pm}}}
\def \Rm {\ensuremath{\rm{Rm}}}

\def\beg{\begin{equation}}
\def\ende{\end{equation}}
\newcommand{\gsim}{\lower.7ex\hbox{$\;\stackrel{\textstyle>}{\sim}\;$}}
\newcommand{\lsim}{\lower.7ex\hbox{$\;\stackrel{\textstyle<}{\sim}\;$}}
\renewcommand{\vec}[1]{\mbox{\boldmath $#1$}}
\def\curl{{\rm curl}} 
\def\Om{{\it \Omega}}

\def\etaT{\eta_{\rm t}}
\def\meta{{\bar \eta}}
\def\etarms{\eta_{\rm rms}}
\def\urms{u_{\rm rms}}

\def\ara\&a{ Ann. Rev. Astronomy Astrophysics}

\shorttitle{Fluids with fluctuating electric conductivity}
\shortauthor{G. R\"udiger,  M. K\"uker and  P. J. K\"apyl\"a}

\title{Electrodynamics of turbulent fluids with  fluctuating electric conductivity
\\   
}

\author{G. R\"udiger\aff{1,2}
  \corresp{\email{gruediger@aip.de}}, 
 M. K\"uker\aff{1}
 \and  P. J. K\"apyl\"a\aff{3,4} 
 }

\affiliation{\aff{1}Leibniz-Institut f\"ur Astrophysik Potsdam, An der Sternwarte 16, D-14482 Potsdam, Germany
\aff{2}University of Potsdam, Institute of Physics and Astronomy, Karl-Liebknecht-Str. 24-25, 14476 Potsdam, Germany
\aff{3}Institut f\"ur Astrophysik, Georg-August-Universit\"at G\"ottingen, D-37077 G\"ottingen, Germany
\aff{4}ReSoLVE Centre of Excellence, Department of Computer Science, P.O. Box 15400, FI-00076 Aalto, Finland
}

\begin{document}

\maketitle

\begin{abstract}
Consequences of fluctuating microscopic conductivity in mean-field electrodynamics of turbulent fluids are formulated and discussed.  
If the
conductivity fluctuations are assumed  to be uncorrelated with the
 velocity fluctuations then only the turbulence-originated magnetic diffusivity
of the fluid is reduced and the decay time of a large-scale magnetic
field or the cycle times of oscillating turbulent dynamo models are increased. If, however,  the fluctuations
of conductivity and  flow in a certain well-defined direction
are correlated, an
additional diamagnetic pumping effect results transporting magnetic
field in opposite direction to the diffusivity flux vector $\langle
\eta'\vec{u}'\rangle$. In the presence of global rotation even for  homogeneous 
turbulence fields  a new $\alpha$ effect  appears.
If the characteristic values of the outer core of the Earth or
the solar convection zone are applied, the dynamo number of the 
$\alpha$ effect does not reach
supercritical values to operate as an $\alpha^2$-dynamo but   oscillating $\alpha\Om$-dynamos with differential rotation are not excluded.
\end{abstract}
\keywords  {Astrophysical plasma -- dynamo theory}
\section{Introduction}
The electromotive force (EMF) $\vec{u}\times \vec{B}$ is the only
nonlinear term in the induction equation on which the present-day
mean-field electrodynamics { is based on}. It is  the only
nonlinear term
in this equation if the microscopic magnetic diffusivity $\eta$ in the
fluid is uniform. This, however, is not necessarily true. If by any
reason the electric conductivity fluctuates around a certain average  value
then the local diffusivity { fluctuates around its basal value} so that the
effective decay time of a large-scale electric current is 
changed. Below we shall demonstrate  this phenomenon -- which reduces the effective eddy diffusivity of
a turbulence field \citep{KR73} --  
 also with { nonlinear simulations}.

In convection-driven turbulent fields temperature fluctuations should 
produce electric-conductivity fluctuations which are correlated with
the vertical component of the flow field. In this case even  a turbulent  diffusivity
flux vector $\langle\eta'\vec{u}'\rangle$ occurs 
which in connection with the large-scale field and/or the large-scale
electric current may form new terms in the mean-field induction
equation. \cite{PA16} assumed that a new sort of alpha effect
 arises in such systems.  Our considerations  confirm the existence of an alpha effect
but only in the presence of global rotation. Without rotation the conductivity fluctuations will (only) lead to a reduction of the eddy diffusivity and -- if correlated with one of the velocity components -- to a new diamagnetic pumping term.

\section{The Equations}
\label{sec:Equs}
{ The problem is mainly  described by the induction equation}
\begin{eqnarray}
 \frac{\partial \vec{B}}{\partial t}&=& {\textrm{curl}} \Big(\vec{u} \times \vec{B} -
 \eta\ {\textrm{curl}} \vec{B}   \Big)  
   \label{mhd1}
\end{eqnarray}
with $  {\textrm{div}}\ \vec{B} = 0$ and  $  {\textrm{div}}\ \vec{u}  = 0$ for an incompressible fluid.
Here $\vec{u}$ is the velocity, $\vec{B}$ { is} the magnetic field
vector and $\eta$  the magnetic diffusivity. { We 
consider} a turbulent fluid with
$\vec{u}= \bar{\vec{u}}+ \vec{u}' $ and with a fluctuating magnetic
diffusivity ${\eta}= \meta+ {\eta}' $. For the expectation values of
the perturbations we shall use the notations 
$\urms={\langle {\vec{u}'}^2 \rangle}$$^{1/2}$ and $\etarms={\langle {\eta'}^2 \rangle}$$^{1/2}$. 
{ Large-scale observables { (mean values)} are marked with overbars
  while brackets are used for the correlations { of fluctuations}.
  For finite fluctuations
the high-conductivity limit $\meta\to 0$ is not allowed. The
fluctuations $\vec{u}'$ and $\eta'$ may be correlated so that
a turbulence-originated diffusivity flux 
\beg
\vec{U}=\langle \eta' \vec{u}'\rangle
\label{U}
\ende
 forms a vector which is polar  by definition. The existence of this vector is obvious  for thermal convection, when both the velocity field and the electric conductivity are  due to temperature fluctuations. The correlation (\ref{U}) can be understood as  transport of magnetic diffusivity in a preferred  direction. Also the magnetic field will fluctuate, i.e. $\vec{B}= \bar{\vec{B}}+ \vec{B}' $. The magnetic fluctuation $\vec{B}'$ fulfills a nonlinear  induction equation which follows from (\ref{mhd1}). We shall  only discuss its linear
version
\begin{eqnarray}
 \frac{\partial \vec{B}'}{\partial t}&=& {\textrm{curl}} \Big(\vec{u}' \times \bar{\vec{B}} -
 \meta\ {\textrm{curl}} \vec{B}'  -  \eta'\ {\textrm{curl}} \bar{\vec{B}}  \Big)  
   \label{mhd11}
\end{eqnarray}
{in the analytical theory of driven turbulence \citep{KR80}. The
  results of the calculations within the quasilinear  First Order Smoothing
  Approximation (FOSA) will also be probed by { targeted} numerical
  simulations with { well-established} nonlinear MHD codes. }

If the
fluctuations are known in their dependences on the magnetic background field and rotation then the turbulence-originated electromotive force (EMF) ${\vec{\cal
    E}}=\langle \vec{u}'\times \vec{B}' \rangle$ and the  diffusivity-current correlation
\beg
\vec{{\cal J}}=-\langle \eta' \curl
\vec{B}'\rangle
\label{corr}
\ende
can be formed which enter  the induction equation for the large-scale field via
\beg
 \frac{\partial \bar{\vec{B}}}{\partial t}= {\textrm{curl}} \Big(\vec{\cal E} + \vec{\cal J}
- \meta\ {\textrm{curl}} \bar{\vec{B}}   \Big).  
   \label{mhd0}
\ende
To find the influence of a large-scale field and/or its  gradients on the  EMF
${\vec{\cal E}}$ at linear order it is enough to solve
the induction equation (\ref{mhd11})
where the inhomogeneous large-scale  magnetic field may be  written in the form
$
\bar B_j= B_{jp} x_p
$
with ${B}_{jp}\equiv  {\bar B}_{j,p}$. { Without any loss of generality  the  coordinate ${\vec  x}=0$  defines the point where the background field vanishes}. We also  note  that the global
rotation here  only appears in the Navier-Stokes equation for the velocity fluctuation
 which remains homogeneous if only expressions linear in $B_{jp}$
are envisaged. One can thus work with 
\begin{eqnarray}
u'_i (\vec{x},t)&=& \int\!\!\!\!\int \hat u_i (\vec{k},\omega)
 e^{{\rm i}({\bf k}{\bf x}-\omega t)} {\rm d}\vec{k} \, {\rm d}\omega, 
\nonumber\\
B'_i (\vec{x},t)& =& \int\!\!\!\!\int (\hat B_i (\vec{k},\omega)
 + x_l \hat B_{il} (\vec{k},\omega)) e^{{\rm i}({\bf k}{\bf x} -\omega t)}
 {\rm d} \vec{k} \, {\rm d}\omega.
\label{mhd2}
\end{eqnarray}
The result is 
\beg
\hat B_i =  \frac{{\rm i}x_l k_j B_{jl}}{-{\rm i} \omega
 + \meta k^2} \hat u_i  -\frac{B_{ij} + \frac{2\meta k_l k_m B_{lm} \delta_{ij}}
{-{\rm i}\omega + \meta k^2}}{-{\rm i} \omega + \meta k^2} \hat u_j
+  \frac{{\rm i}  k_j (B_{ij}-B_{ji})}{-{\rm i} \omega
 + \meta k^2}\hat\eta 
\label{mhd3}
\ende
as the spectral component of the magnetic fluctuations \citep{RK13}. The first two terms on the r.h.s. of this equation  describe the interaction of the turbulence with the large-scale magnetic field and its gradients. 
Under the assumption that the large-scale  field $\bar{\vec{B}}$   varies slowly 
 in space and time, the electromotive force can be written as
\begin{eqnarray}
\vec{{\cal E}} = \alpha \circ \vec{\bar{B}} - \etaT \curl \vec{\bar{B}},
\label{alpha}
\end{eqnarray}
where the tensor $\alpha$ and the coefficient $\etaT$ 
represent the $\alpha$ effect and the  turbulent magnetic diffusivity.  

The  last term in (\ref{mhd3}) directs the influence of the fluctuating diffusivity. It  leads to an EMF of 
\beg
{\cal E}_i=  \epsilon_{iqp} \int\!\!\!\!\int \frac{{\rm i} k_j  \hat U_q}{{-{\rm i} \omega + \meta k^2} } {\rm d}\vec{k}\, {\rm d} 
\omega  \left(B_{pj}-B_{jp} \right),
\label{mhd4}
\ende
where $\hat{\vec{U}}$ is the Fourier transform of the diffusivity-velocity correlation
$\vec{U}$ which itself is  a polar vector. 
The spectral vector of the correlation (\ref{U}) can in full generality  be written as
\beg
{\hat U}_i=u_1 \left[ g_i-\frac{(\vec{g}\vec{k})k_i}{k^2} \right] + u_2 {\rm i} \epsilon_{ijk} k_j g_k.
\label{Ucorr}
\ende
The vector $\vec g$  gives the unit vector of the coordinate in which
direction the correlation between velocity and diffusivity is
non-vanishing. The expression (\ref{Ucorr}) must be odd in $\vec{g}$ and the real
part must be even in the wave number $\vec{k}$. The quantity $u_1$
reflects the correlation of the velocity component $\vec{g}\vec{u}'$
with $\eta'$. The second term in (\ref{Ucorr}) contains a correlation
of diffusivity and vorticity where $u_2$ must be a pseudoscalar. Equations (\ref{mhd4}) and (\ref{Ucorr})
lead to
\beg
\vec{\cal E}=   \int\!\!\!\!\int \frac{\meta k^4   u_2}{{ \omega^2 + \meta^2 k^4} } {\rm d}\vec{k}\, {\rm d} \omega\ \ \   \vec{g}\times \vec{\bar J}
\label{mhd44}
\ende
with $\vec{\bar J}=\curl \bar{\vec{B}}$ and $(\vec{g}\times \vec{\bar J})_i=- (\vec {g}\cdot\nabla)\bar{B}_i  + \ \nabla_i...$
where the latter symbol  represents a gradient which does not play a role in the induction equation. We note that the non-potential  term  only exists if the  magnetic field  $\bar{\vec B}$   depends on the coordinate along  $\vec g$. 

\subsection{The  diffusivity-current correlation}
{ The diffusivity-current correlation $\vec{{\cal J}}$ from (\ref{corr})  is now analyzed
in detail}. Fourier transformed fluctuations of the electric current are
\beg
 \curl_i \hat{ \vec{B}} =- \frac{ \epsilon_{isp}k_jk_s}{-{\rm i} \omega+ \meta k^2} \left(   \hat{u}_pB_j + \hat \eta 
 (B_{pj}-B_{jp})\right).
\label{mhd5}
\ende
Multiplication with the (negative) Fourier transform of the diffusivity fluctuation, $\hat\eta$, leads to 
\beg
 \hat{\cal J}_i = \frac{ \epsilon_{isp}k_jk_s}{-{\rm i} \omega+ \meta k^2} \left(\hat{U}_p B_j + {\hat V} 
 (B_{pj}-B_{jp})\right).
\label{mhd6}
\ende
 $\hat V$ { is here  the spectral function of the autocorrelation function}
$
V=\langle \eta'(\vec{x},t)\eta'(\vec{x}+\vec{\xi}, t+\tau)\rangle
$
of the diffusivity fluctuations. Equations  (\ref{Ucorr})  and (\ref{mhd6}) provide 
$
 \vec{{\cal J}} =-\gamma \ \vec{g}\times \bar{\vec{B}}
$
with
\beg
 \gamma= \frac{1}{3} \int\!\!\!\!\int \frac{\meta k^4  u_1}{{ \omega^2 + \meta^2 k^4} } {\rm d}\vec{k}\, {\rm d} \omega,
\label{gamma}
\ende
representing  a turbulent transport of the magnetic background field (`pumping') anti-parallel to  $\vec g$. For positive $u_1$ (i.e. for positive correlation of $\eta'$ and $u'_z$) the pumping goes downwards as $\vec g$ is the vertical unit vector. 
We note  that formally the integral in { (\ref{gamma})} also exists in the
high-conductivity limit $\meta\to 0$ so that for small $\meta$ it does not depend on the
magnetic Reynolds number 
\beg
{\Rm}=\frac{\urms\ell}{\meta}
\label{rm}
\ende
(with $\ell$ as the correlation length) for large $\Rm$. In this limit
$\gamma$ is linear in the correlation function $u_1$. For small $\Rm$
the integral in { (\ref{gamma})} linearly runs with $\Rm$ which follows
after application of the extremely steep correlation function
$\delta(\omega)$ as a proxy of  the low-conductivity limit.

On the other hand, { the term with  $\hat V$ in Eq. (\ref{mhd6}) leads to}
\beg
\vec{\cal J}= .... +  \frac{2}{3} 
\int\!\!\!\!\int  \frac{k^2 {\hat V}}{-{\rm i} \omega + \meta k^2 } 
{\rm d}\vec{k}\, {\rm d}\omega \ \curl \bar{\vec{B}},
\label{mhd7}
\ende
which provides an extra contribution to the magnetic field
dissipation. The question is whether this term reduces or enhances the
{  eddy diffusivity $\etaT$ which is due to the turbulence without $\eta$-fluctuations.} 
For homogeneous turbulence one finds from (\ref{mhd3})
\beg
{\cal E}_i= - \epsilon_{ijp} \int\!\!\!\!\int \left(B_{pn} \hat Q_{jn}
 + \frac{2\meta k_l k_m}{-{\rm i} \omega + \meta k^2} B_{lm} \hat Q_{jp}\right)
 \frac{{\rm d}\vec{k}\, {\rm d} 
\omega}{-{\rm i} \omega + \meta k^2} .
\label{mhd8}
\ende
The spectral tensor $\hat Q_{ij}$ for isotropic turbulence   is
\beg 
\hat Q_{ij}(\vec{k}, \omega)=\frac{E(k,\omega)}{16\pi k^2}
 \left(\delta_{ij}-\frac{k_i 
k_j}{k^2}\right) - {\rm i} \epsilon_{ijk} \, k_k \, H(k,\omega), 
\label{qu} 
\ende 
where the positive-definite spectrum $E$ gives the energy
\beg 
u^2_{\rm rms} =
\int\limits_0^\infty\!\!\int\limits_0^\infty E(k,\omega) \ 
{\rm d}k \ {\rm d}\omega 
\label{intensity} 
\ende 
and $H$ is  the helical part of the 
turbulence field. From (\ref{mhd8})  $\vec{\cal E}=- \etaT\ \curl \bar{\vec{B}}$ with the eddy diffusivity 
\beg
\etaT=\frac{1}{24\pi} 
\int\!\!\!\!\int  \frac{\meta  E}{ \omega^2 + \meta^2 k^4 } 
{\rm d}\vec{k}\, {\rm d}\omega  
\label{mhd9}
\ende
results  with $\etaT>0$.

For the sum of the  turbulence-originated terms in Eq. (\ref{mhd0}) one    obtains
\begin{eqnarray}
\vec{\cal E}+\vec{\cal J}=  -\left(\frac{1}{24\pi} 
\int\!\!\!\!\int  \frac{\meta  E}{ \omega^2 + \meta^2 k^4 } 
{\rm d}\vec{k}\, {\rm d}\omega -
\frac{2}{3} 
\int\!\!\!\!\int  \frac{\meta k^4 {\hat V}}{ \omega^2 + \meta^2 k^4 } 
{\rm d}\vec{k}\, {\rm d}\omega \right)\ \curl \bar{\vec{B}}-\nonumber\\
- \frac{1}{3} \int\!\!\!\!\int \frac{\meta k^4  u_1}{{ \omega^2 + \meta^2 k^4} } {\rm d}\vec{k}\, {\rm d} \omega  \ \ \  \vec{g}\times \bar{\vec{B}},
\label{mhd10}
\end{eqnarray}
indicating   the total turbulent diffusivity as {\em reduced} by the conductivity fluctuations.  On the other hand, the pumping  term in the second line of this equations only exists if these conductivity fluctuations are correlated with the flow component in a preferred direction within the fluid.
All  terms in Eq. (\ref{mhd10}) also exist in the high-conductivity limit, $\meta\to 0$.

The modified eddy diffusivity is  
\begin{eqnarray}
\frac{\etaT^{\rm eff}}{\meta}=  \frac{1}{24\pi} 
\int\!\!\!\!\int  \frac{ E}{ \omega^2 + \meta^2 k^4 } 
{\rm d}\vec{k}\, {\rm d}\omega -
\frac{2}{3} 
\int\!\!\!\!\int  \frac{ k^4 {\hat V}}{ \omega^2 + \meta^2 k^4 } 
{\rm d}\vec{k}\, {\rm d}\omega 
\label{etat}
\end{eqnarray}
\citep[see][]{KR73}. For large $\Rm$ both terms run
linearly with $\Rm$ while for small $\Rm$ both terms  formally run with
$\Rm^2$. If the second expression is considered as function of $\eta_{\rm rms}/\meta$ then it runs with $1/\Rm$
for large $\Rm$ and with $\Rm^0$ for small $\Rm$. 
As it should, the reduction of the eddy diffusivity by conductivity fluctuations   disappears  in the high-conductivity limit.

Discussing possible dynamo effects in hot Jupiter atmospheres  \cite{R17}  considered variable molecular diffusivities which form patterns in the vertical direction {\em and} the horizontal plane. In the horizontal plane the  quasi-twodimensional velocity field existed without being correlated with the diffusivity.  In consequence, the effective magnetic-diffusivity    is also  reduced as in    (\ref{etat}) but the pumping term (\ref{gamma}) does not appear.

\subsection{Direct numerical simulations}
{To test  theoretical predictions, we run fully nonlinear numerical
simulations with the {\sc Pencil
  Code}\footnote{https://github.com/pencil-code/}. We solved the
equations of compressible magnetohydrodynamics
\begin{eqnarray}
\hspace{1cm} \frac{\partial \vec{A}}{\partial t} = \vec{u}\times \vec{B} - (\meta + \eta')\mu_0 \vec{J} + \vec{\cal E}_0, \label{equ:induc}
\end{eqnarray}
\begin{eqnarray}
\frac{D \ln \rho}{Dt} = -{\textrm{div}}\ \vec{u},\ \ \ \ \ \ \ \ \ \ \ \ \  
\frac{D \vec{u}}{Dt} = - c_{\rm s}^2 \ln \rho + \vec{F}^{\rm visc} + \vec{F}^{\rm force},
\end{eqnarray}
where $\vec{A}$ is the magnetic vector potential and $\vec{B} =\curl \vec{A}$ is the magnetic field, $\vec{J} =
\mu_0^{-1} \curl \vec{B}$ is the current density $D/Dt =
\partial/\partial t + \vec{u} \cdot \vec{\nabla}$ is the advective
time derivative, $\rho$ is the density, and $c_{\rm s}$ is the
constant speed of sound. 
The last term  of Eq.~(\ref{equ:induc}) describes an imposed
EMF
$
\vec{\cal E}_0 = \hat{\cal E}_0 \sin (k_1 x) \hat{\vec{e}}_z,
$
that is used to introduce a large-scale magnetic field $\bar{B}_y(x)$ in the
system. Furthermore, the
fluctuating component of the magnetic diffusivity is given by
$
\eta' = c_u u_z,
$
where $c_u$ is used to control the strength of the correlation. We use
$\eta_{\rm rms}=c_u u_{z, {\rm rms}}$ 
to quantify the
amplitude of the fluctuating part of the diffusivity.

The viscous force is given by the standard expression
\begin{eqnarray}
\vec{F}^{\rm visc} = \nu \left(\nabla^2\vec{u} + {1 \over 3} \vec{\nabla}\vec{\nabla}\vec{\cdot} \vec{u} 
\right),
\end{eqnarray}
where $\nu$ is the kinematic viscosity.
The fluid is forced with an
external body force
$
\vec{F}^{\rm force}(\vec{x},t) = Re\{N \vec{f}_{\vec{k}(t)} \exp[{\rm i} \vec{k}(t)\vec{\cdot} \vec{x} - {\rm i}\phi(t)]\},
$
where $\vec{x}$ is the position vector, $N= f_0 c_{\rm s} (k c_{\rm
  s}/\delta t)^{1/2}$ is a normalization factor where $f_0$ is the
non-dimensional amplitude, $k=|\vec{k}|$, $\delta t$ is
the length of the time step, and $-\pi < \phi(t) < \pi$ is a random
delta--correlated phase. The vector $\vec{f}_{\vec{k}}$ describes
non-helical transversal waves.

The simulation domain is a fully periodic cube with volume
$(2\pi)^3$. The units of length and time are
$
[x] = k_1^{-1}, [t]=(c_{\rm s} k_1)^{-1}$ 
where $k_1$ is the wave number corresponding to the system size. The simulations
are characterized by the  magnetic Reynolds number (\ref{rm})
with $u_{\rm rms}$ volume averaged { and $\ell=(k_{\rm f})^{-1}$}.  
The flows under
consideration are weakly compressible with Mach number ${\rm
  Ma}=u_{\rm rms}/c_{\rm s}\approx0.1$. All of the simulations use
$k_{\rm f}/k_1 = 30$ and a grid resolution of $288^3$.

We first run the simulations with each ${\rm Rm}$ with $\eta'=0$
sufficiently long that a stationary large-scale magnetic field
$\bar{B}_y(x)$ due to the imposed $\vec{\cal E}_0$ is established. The
amplitude of the resulting magnetic field is typically of the order of
$10^{-3}$ of equipartition strength such that its influence on the
flow is negligible. Then we branch new simulations from snapshots of
these runs with different levels of diffusivity fluctuations $\eta'$
and switch off the imposed EMF, i.e.\ $\vec{\cal E}_0=0$. Without the
EMF the large-scale magnetic field decays. 
 Measuring the decay rate of the magnetic field,  the effective turbulent diffusion can be
computed. At least five decay experiments with each value of ${\Rm}$ and $\eta'$ were made and the averaged decay rate  was used in the
computation of $\etaT^{\rm eff}/\meta$. The error bars in Fig.~\ref{fig1}
indicate the standard deviation divided by the square root of the
number of experiments.

Going back to Eq.~(\ref{etat}), we note that if the second expression
is considered  then its argument  $(\eta_{\rm rms}/\meta)^2$ must be multiplied  with
$1/\Rm$ for large $\Rm$ and with $\Rm^0$ for small $\Rm$. It is thus clear that the 
diffusivity-reduction by conductivity fluctuations
disappears in the high-conductivity limit  which is  confirmed by the
numerical results, { see the left panel of Fig.~\ref{fig1}}.
{ The { right} panel { of Fig.~\ref{fig1}} shows the
  numerical results for the ratio $\etaT^{\rm eff}/\etaT$ of the terms
  in (\ref{etat}) which, of course, is unity for vanishing
  $\etarms$. It is also unity for large $\Rm$ as the
  $\eta$-fluctuation-induced second term in (\ref{etat}) vanishes with
  $1/\Rm$. Its role, however, becomes more important for small
  $\Rm$. In this case the first term  looses its
  dominance
and the total diffusivity $\etaT^{\rm eff}$ is reduced. If the numbers
of the { right} plot of Fig.~\ref{fig1} { are} multiplied with $\etaT/\meta$
then the { left} plot results where the total magnetic diffusivity
normalized with the microscopic value $\bar{\eta}$ is given. The
influence of the conductivity fluctuations vanishes for large $\Rm$
while the fluctuations provide smaller effective diffusivities
$\etaT^{\rm eff}$ so that the cycle frequencies of oscillating dynamo
models are reduced \citep{R72}, also  characteristic growth and decay  times  become longer .
}
\begin{figure}
  \centerline{
  \includegraphics[width=0.5\textwidth]{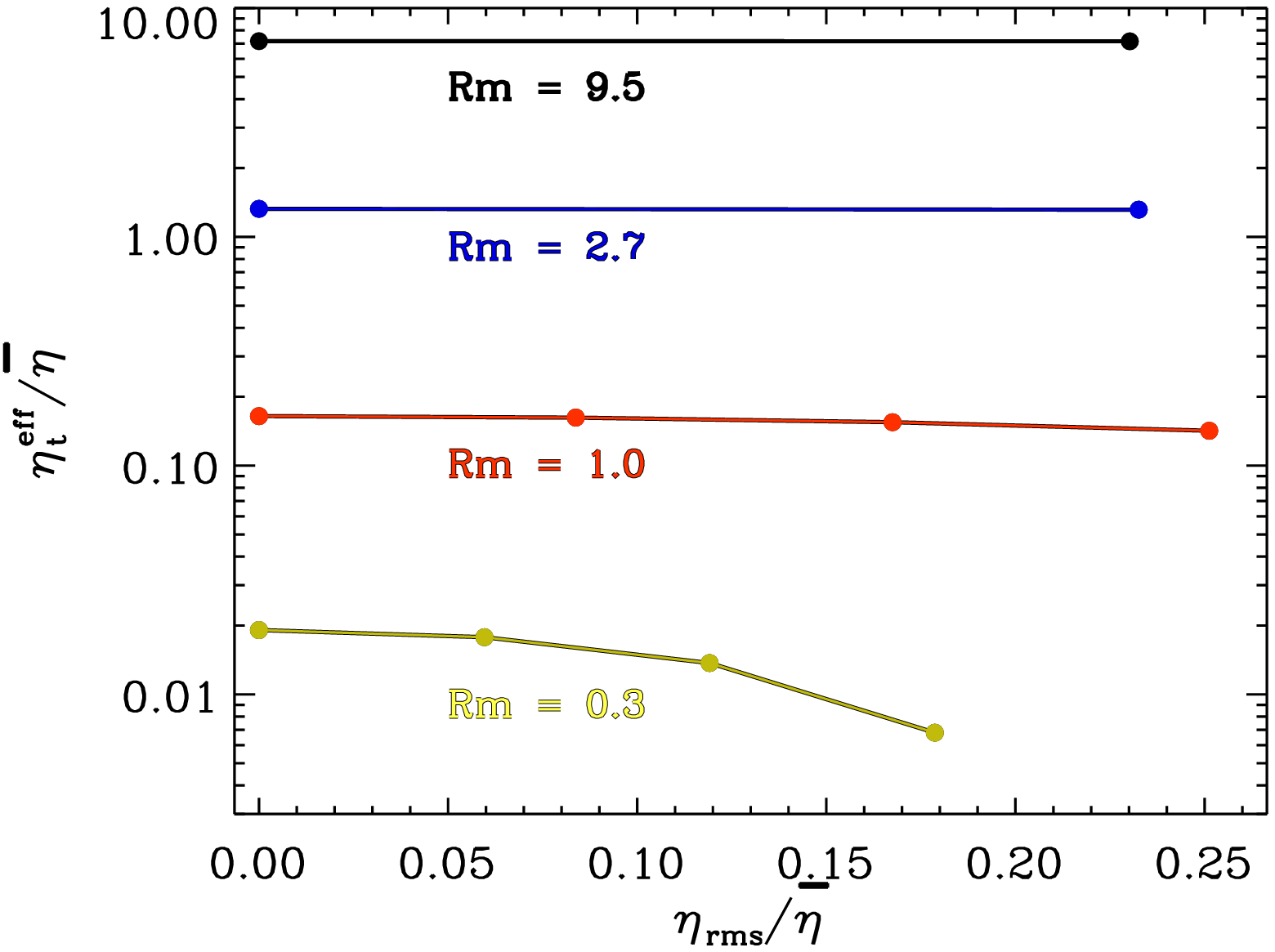}
  \includegraphics[width=0.5\textwidth]{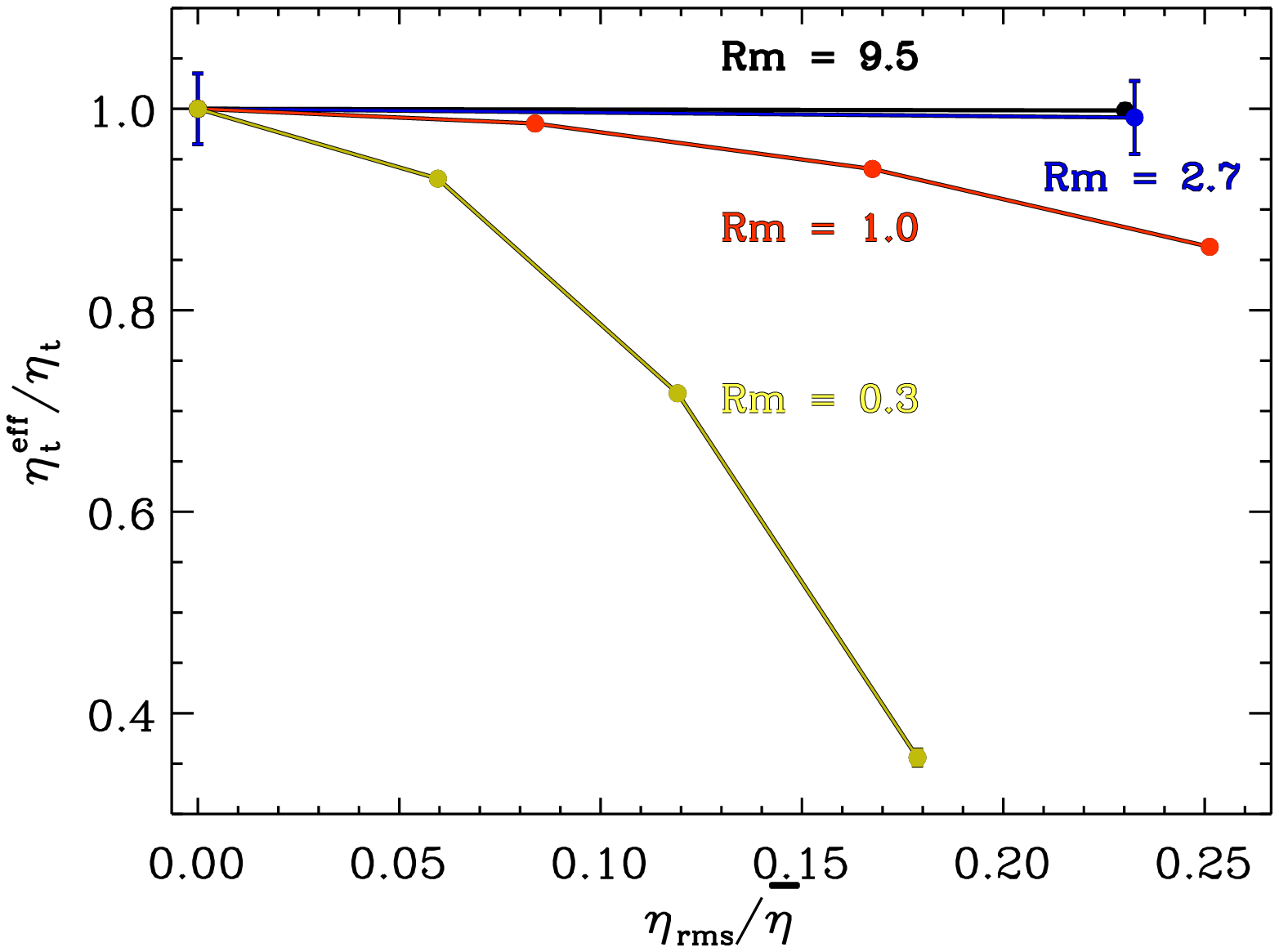} }
  \caption{{ The specific diffusivity $\etaT^{\rm eff}/\meta$
      (left) and the eddy diffusivity ratio $\etaT^{\rm eff}/\etaT$
      (right)} as functions of the normalized diffusivity fluctuation
    $\eta_{\rm rms}/\meta$. In the high-conductivity limit ($\Rm\gg
    1$) the influence of the conductivity fluctuations disappears.}
\label{fig1}
\end{figure}

\section{Alpha effect} 
All   turbulent flows which are known to possess an alpha effect  are helical due to an inhomogeneity in the
rotating turbulence field subject to  the influence of a density and/or turbulence-intensity stratification\footnote{Also global shear needs a stratification to develop $\alpha$ effect \citep{RB14}.}.  The  product $\vec{g}\cdot \vec{\Om}$ forms the
pseudo-scalar on which  the pseudo-tensor $\alpha$ in the relation (\ref{alpha}) bases. 
However, the  turbulence model considered in this paper is homogeneous and anisotropic.
As the anisotropy is only implicit, it is not trivial whether the
 influence of global rotation will lead to an alpha effect or not.
\subsection{Quasilinear approximation}
We start with Eq. (\ref{mhd5}) and include  the influence of rotation by the transformation
$\hat u_p= D_{pq} \hat u_q$ with the rotation operator
\beg
D_{ij}= \delta_{ij}
+ \frac{(2\vec{k}\cdot\vec{\Om}/k)}
{-{\rm i}\omega + \nu k^2} 
\epsilon_{ijp} \frac{k_p}{k}
\label{D}
\ende
in linear approximation \citep{KP94}. { The second term gives the
  influence of the basic rotation in the Fourier representation. As it
  should, it is even in the wave number and odd in the angular
  velocity. The { Levi--Civita} tensor ensures that the term
  { is invariant with respect to the transformation} of the
  coordinate system}. It follows { that}
$
 \hat {\cal J}_i ={ { \epsilon_{isp}k_jk_s D_{pq}\hat U_q}B_j}/({-{\rm i} \omega+ \meta k^2})  + ....,
$
 and finally 
 \beg
\vec{\cal J}= -\gamma\ \vec{g}\times \vec{B} +\alpha \left( 4 (\vec B\cdot\vec\Om)\vec g - (\vec g\cdot \vec B)\vec{\Om} -(\vec{g}\cdot\vec{\Om})\vec{B}  \right),
\label{J2}
\ende
 where $\gamma$ is given by Eq. (\ref{gamma}) and for the coefficient $\alpha$ (related  but not identical to the tensor $\alpha$ in (\ref{alpha})) one finds
 \beg
 \alpha= \frac{2}{15} \int\!\!\!\!\int \frac{(\nu\meta k^4-\omega^2) k^2  u_1}{(\omega^2 + \meta^2 k^4) (\omega^2 + \nu^2 k^4)} {\rm d}\vec{k}\, {\rm d} \omega.
\label{intalpha}
\ende
For $\nu=\meta$ and for frequency spectra which monotonously decrease for increasing $\omega$ the frequency integral in (\ref{intalpha}) has the same sign as $u_1$ while it vanishes a for a spectrum (``white-noise'')  { which does not depend on the frequency $\omega$. Correlations of a  white-noise spectrum possess zero correlation times so that indeed the rotational influence  should vanish.}
The $\alpha$ effect after (\ref{J2}) is highly anisotropic, its last
term is the rotation-induced standard $\alpha$ expression. 

Both
quantities $\alpha$ and $\gamma$ are linearly running with the ratio
$\eta_{\rm rms}/\meta$. In the low-conductivity limit ($\Rm<1$) { they are}
\beg
 \frac{\gamma}{\urms}\simeq \frac{\eta_{\rm rms}}{\meta}\ \ \ \ \ \ \ \ \ \ \ \ \ \ \ \ \ \ \ \ \ \ \    \frac{\gamma}{\alpha\Om}\simeq \frac{1}{\Rm\ (\tau_{\rm corr}\Om)}
\label{alpha1}
\ende
 while in the high-conductivity limit ($\Rm>1$)
   \beg
 \frac{\gamma}{\urms}\simeq \frac{\eta_{\rm rms}}{\meta}\frac{1}{\Rm}\ \ \ \ \ \ \ \ \ \ \ \ \ \ \ \ \ \ \ \ \ \ \    \frac{\gamma}{\alpha\Om}\simeq \frac{1} {\tau_{\rm corr}\Om}
\label{alpha2}
\ende
-- both relations for the $\alpha$ terms taken for all $\Pm=\nu/\meta\leq 1$.  The
$\alpha$ effect always needs rotation; both the given  coefficients are small.
The dimensionless ratio $\hat\gamma=\gamma/\alpha\Om$ of the pumping term $\gamma$ and the  $\alpha$
 effect indicates the ratio of off-diagonal and diagonal elements in the alpha tensor.
For $\hat\gamma>1$ dynamo operation can highly be disturbed.
For a standard disk dynamo \cite{RE93} demonstrated with numerical
simulations that large values of $|\hat\gamma|$ suppress
the dynamo action. In spherical dynamo models the $\vec{\gamma}$ term
plays the role of an upward buoyancy \citep{MT90} or even a strong
downward turbulent pumping \citep{BM92}. In order to be relevant for
dynamo excitation the $\alpha$ effect should numerically exceed the
value $\gamma$ of the pumping term.
As the pumping effect { exists even} for $\Om=0$, the ratio
$\hat\gamma$ should decrease for faster rotation.
With extensive numerical simulations  \cite{GZ08} derived values of order unity for interstellar turbulence driven by collective  supernova explosions. For rotating magnetoconvection  
\cite{OS01,OS02} also found $\hat\gamma\simeq 1$  where both $\alpha$ and $\gamma$  reached about 10\% of the rms value of the convective velocity.
 In their simulations of turbulent magnetoconvection also \cite{KK09} reached typical values of order unity for  $\hat\gamma$.
\subsection{Turbulent transport of electric current}  
{ We shall demonstrate why} the existence of a diamagnetic pumping and
an $\alpha$ effect for rotating but unstratified fluids with
fluctuating diffusivity (in a fixed direction) is not too
surprising. We start with the flow-current correlation $\langle
\vec{u}' \cdot \curl \vec{B}'\rangle$ describing a turbulent transport
of electric current fluctuations which after (\ref{mhd5}) for
non-rotating turbulence certainly vanishes. This is not true
for rotating turbulence as $\langle \vec{u}' \cdot \curl
\vec{B}'\rangle \propto \bar{\vec{B}}\cdot \vec{\Om}$ is a possible
construction for {\em isotropic turbulence} fields which only vanishes
for $\bar{\vec{B}}\perp \vec{\Om}$. Moreover, the tensor $\langle
{u}_i' \ \curl_j \vec{B}'\rangle $ for rotating isotropic turbulence
may be written as
\beg
\langle {u}_i'\  \curl_j \vec{B}'\rangle = \kappa_1 \Om_i {\bar B}_j +\kappa_2 \Om_j {\bar B}_i + \kappa_3 (\vec{\Om}\cdot \vec{\bar B})  \delta_{ij}.
\label{alpha4}
\ende
In opposition to the tensors forming the helicity, the current
helicity and the cross helicity, the tensor (\ref{alpha4}) is not a
pseudo-tensor and there is no reason that the dimensionless
coefficients $\kappa_i$ identically vanish. The correlation
$\langle {u}_r' \ \curl_\phi \vec{B}'\rangle$
describes the up- or downward flow of azimuthal electric current
fluctuations in a rotating magnetized turbulence. { Imagine that
  $u_r'$ is correlated (or anticorrelated) with fluctuations $\eta'$
  of the magnetic diffusivity, i.e. $\langle {u}_r' \ \curl_\phi
  \vec{B}'\rangle\propto \langle \eta' \ \curl_\phi \vec{B}'\rangle$
  {which is  proportional} to ${\cal J}_\phi$. If this quantity occurs for
  rotating turbulence under the influence of an azimuthal magnetic
  background ${\bar B}_\phi$ field then the existence of a new $\alpha $ effect has
  been proven.}
  
The calculation on basis of Eqs. (\ref{mhd3}) and (\ref{D}) for  rotating and magnetized  but otherwise isotropic turbulence  leads to the tensor expression
\beg
\langle {u}_i'\  \curl_j \vec{B}'\rangle = \kappa ( \Om_i {\bar B}_j + \Om_j {\bar B}_i - 4 (\vec{\Om}\cdot {\vec{\bar B}})  \delta_{ij}),
\label{alpha5}
\ende
which is symmetric in its indices.  One finds
\beg
 \kappa= \frac{1}{30} \int\!\!\!\!\int \frac{(\nu\meta k^4-\omega^2) k^2  E}{(\omega^2 + \meta^2 k^4) (\omega^2 + \nu^2 k^4)} {\rm d}{k}\, {\rm d} \omega.
\label{kappa}
\ende
The dimensionless $\kappa$  is almost identical with   the integral (\ref{intalpha}); it is also positive for 
monotonously decreasing frequency spectra (at least for
$\nu=\meta$). { For $\meta\to 0$ one formally finds for the
  integrals $\kappa\simeq {\rm St}^2/15$ where the Strouhal number
  ${\rm St}=u_{\rm rms}\tau_{\rm corr}/\ell$, { with $\ell$ being
    the correlation length}.}
In the low-conductivity limit it runs with 
$\Rm^2$.

On the other hand, without rotation the tensor (\ref{alpha5}) of the homogeneous turbulence  can simply be written as 
$
\langle {u}_i'\  \curl_j \vec{B}'\rangle =  \kappa'  \epsilon_{jik} {\bar B}_k$. As it should, the tensor is invariant against the simultaneous transformation $i\to j$ and $\bar B_k\to -\bar B_k$.
 Then
$
\langle (\vec{g}\cdot \vec{u'})\  \curl \vec{B}'\rangle =  \kappa' \vec{g} \times  \vec{\bar B}
$
for all directions $\vec{g}$, hence
\beg
\langle {u}_r'  \curl_\theta \vec{B'}\rangle = - \kappa' {{\bar B}_\phi}
\label{kadash} 
\ende
for  azimuthal background fields. After the heuristic
replacement of $u_r'$ by $\eta'$, $\kappa'$ in (\ref{kadash})
stands for the new pumping term discussed above. The coefficient
\beg
\kappa'=\frac{1}{15} 
\int\!\!\!\!\int  \frac{\meta k^4  E}{ \omega^2 + \meta^2 k^4 } 
{\rm d}{k}\, {\rm d}\omega,
\label{gamm6}
\ende
which is of the dimension of the inverse of the correlation time,  is positive-definite. 
{ In the formal  limit $\meta\to 0$  the integral yields $\kappa'\simeq (2/15) {\rm St}^2/\tau_{\rm corr}$ { whereas in} the low-conductivity limit it runs with 
$\Rm$.}
We shall further demonstrate by numerical simulations that the
correlations  (\ref{alpha5})  and (\ref{kadash}) { indeed} exist and that the coefficients $\kappa$
and $\kappa'$ are positive.

 \subsection{Rotating magnetoconvection}
A nonlinear numerical simulation with an existing code  demonstrates the existence of the scalar quantities $\kappa'$ and  $\kappa$  and, therefore, the existence of the pumping term (\ref{gamma}) and the new $\alpha$ effect. To this end the correlations $\langle {u}_r'  \curl_\theta \vec{B'}\rangle$  and $\langle {u}_r'  \curl_\phi \vec{B'}\rangle$ are calculated without and with rotation yielding $\kappa'$ and  $\kappa$. As the latter correlation needs global rotation to exist also the $\kappa$ and, therefore, the $\alpha$ effect needs global rotation to exist.

A
convectively unstable Cartesian box penetrated by an azimuthal
magnetic field (fulfilling pseudo-vacuum boundary conditions at top
and bottom of the box) is considered with both density and temperature
stratifications of (only) 10\%.  A detailed description of the
magnetoconvection code has been published earlier \citep{RK16}. The
box is flat: two units in vertical direction and four units in the two
horizontal { directions}, there are $128\times 256\times 256$ grid
points. In code units the molecular diffusivity is $\eta\simeq 6\times
10^{-3}$ and the resulting turbulence intensity $\urms\simeq 0.7$. The
convection cells are characterized by $\tau_{\rm corr}\simeq 0.6$, hence 
$\Rm\lsim 50$. The values are not varied for the various simulation
runs.
After the definitions the magnetic field $B_\phi=1$ would take 40\% of
the equipartition value $B_{\rm eq}=\sqrt{\mu_0\rho}u_{\rm rms}$.
     
The left panel of Fig. \ref{fig3} gives the results of a numerical simulation for a non-rotating box penetrated by an  azimuthal magnetic  field. We find $\kappa'>0$ in accordance with the result (\ref{gamm6}) obtained within  the quasi-linear approximation. If additionally $u_r'$ and $\eta'$ are (say) positively correlated then (\ref{gamm6}) provides positive values of $\gamma$ in accordance to (\ref{gamma}).
{ Multiplication of the numerical result in the left panel of
  { Fig.~\ref{fig3}} with the computed correlation time leads to
  $\tau_{\rm corr}\kappa'\simeq 0.5$ in good agreement with the
  analytical result (\ref{gamm6})}.

From (\ref{alpha5}) also follows that the tensor trace 
$\langle \vec{u}'\cdot \curl \vec{B}'
\rangle=-10\ \kappa\ (\vec{\Om}\cdot {\vec{\bar B}})$ { has a sign
  opposite to that of} the correlation
$\langle (\vec{g}\cdot\vec{u}')\ \curl \vec{B}'\rangle = \kappa
(\vec{g}\cdot\vec{\Om}) \vec{{\bar B}}$, which we now consider for the
hemisphere where $\vec{g}\cdot\vec{\Om}>0$.  One finds that
fluctuations of electric currents in direction of the large-scale
magnetic background field are correlated with the velocity component
$\vec{g}\cdot\vec{u}'$, provided $\vec g$ is not perpendicular to the
rotation axis.
Hence,
\beg
\langle {u}_r' \ \curl_\phi \vec{B}'\rangle = \kappa \cos\theta \Om {\bar B}_\phi,
\label{corrnew}
\ende
 which means  that in a rotating but otherwise isotropic turbulence with an azimuthal background field the
radial flow fluctuations will always be correlated with azimuthal
electric current fluctuations. The correlation (\ref{corrnew}) runs
with $\cos\theta$, it is thus antisymmetric with respect to the
equator and it vanishes there. An upflow motion provides a positive
(negative) azimuthal electric-current fluctuation while a downflow
motion provides a negative (positive) azimuthal electric-current
fluctuation so that the products of $u'_r$ and $ \curl_\phi \vec{B}'$
have the same sign in both cases. Replace now $u'_r$ by $\eta'$ and
the existence of correlations such as $\langle \eta' \curl_\phi {\vec
  B}'\rangle$ becomes obvious in rotating isotropic turbulence fields
magnetized with an azimuthal background field. Just this finding is
formulated by Eqs. (\ref{corr}) and (\ref{J2}). Hence, the
dimensionless coefficient $\kappa$ in (\ref{corrnew}) is a proxy of an $\alpha$ effect which appears when $u_r'$ and $\eta'$ are
correlated or anticorrelated.
\begin{figure}
  \centerline{
  \includegraphics[width=0.45\textwidth]{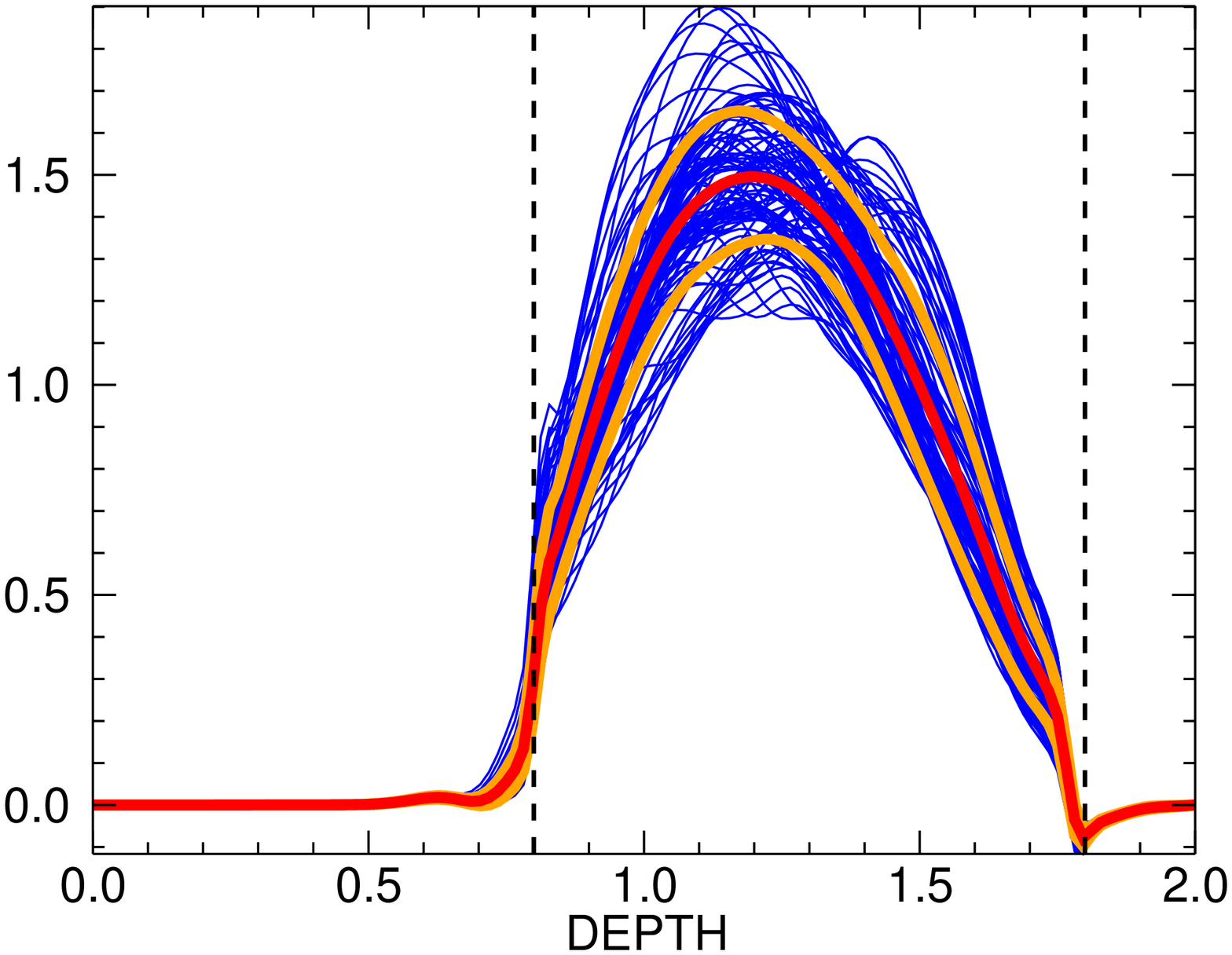}
  \includegraphics[width=0.45\textwidth]{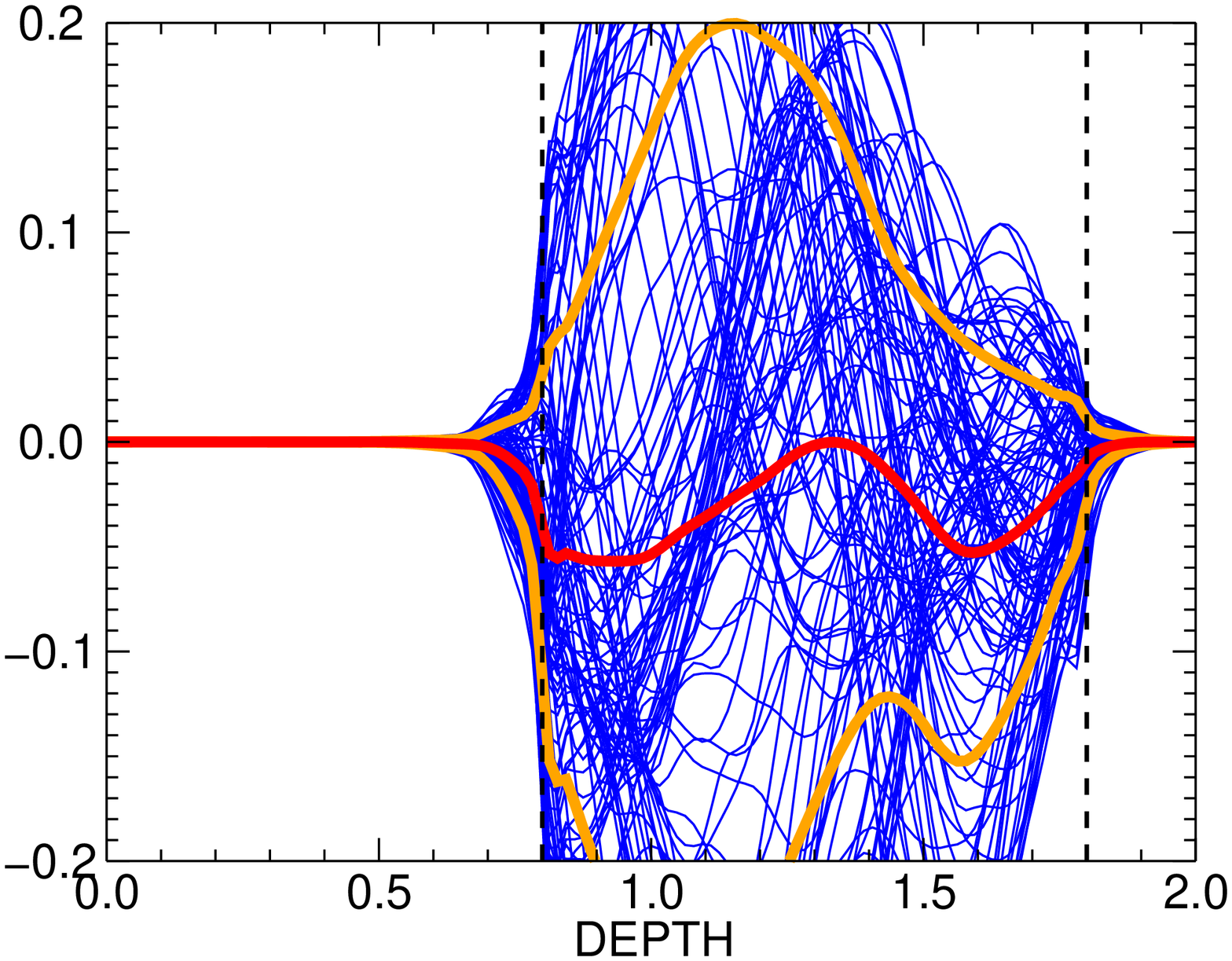}
 }
  \caption{Snapshots of the turbulence-induced coefficients $\kappa'$
    after (\ref{kadash}) (left panel) and the correlation (\ref
{corrnew}) (right panel) from simulations of { non-rotating
    convection} with azimuthal magnetic field.
   The { convectively unstable region} is located between the two vertical
  dashed lines, the red curves denote time averages { and the
    yellow curves characterize the expectation value of the
    fluctuations}. For { non-rotating convection} the correlation $\langle {u}_r'
  \curl_\theta \vec{B'}\rangle$ exists but $\langle
       {u}_r' \ \curl_\phi \vec{B}'\rangle$ vanishes.  $B_\phi=1$, $\Om=0$,
       $\Pm=0.1$. }
\label{fig3}
\end{figure}
\begin{figure}
   \centerline{
   \includegraphics[width=0.34\textwidth]{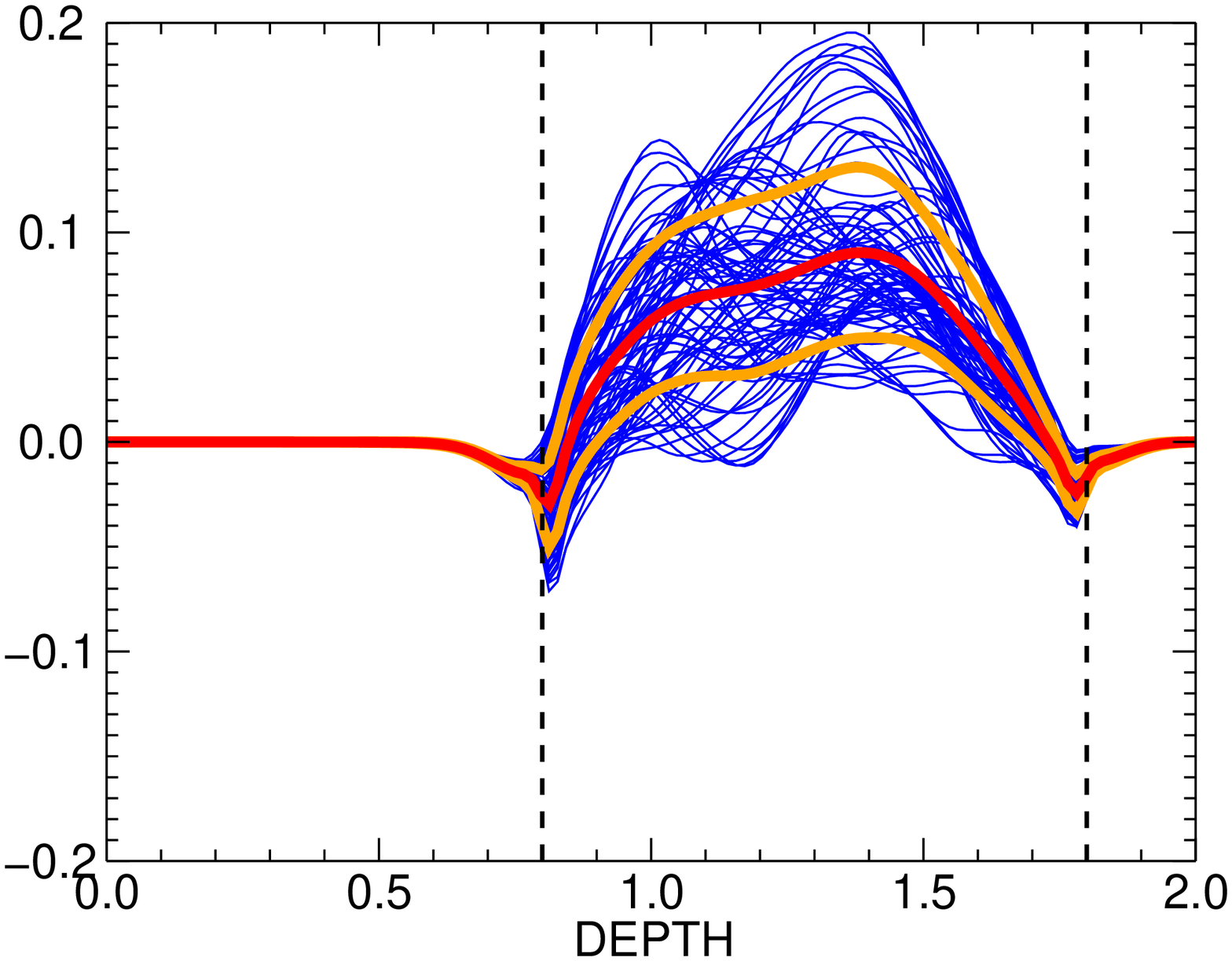}
   \includegraphics[width=0.34\textwidth]{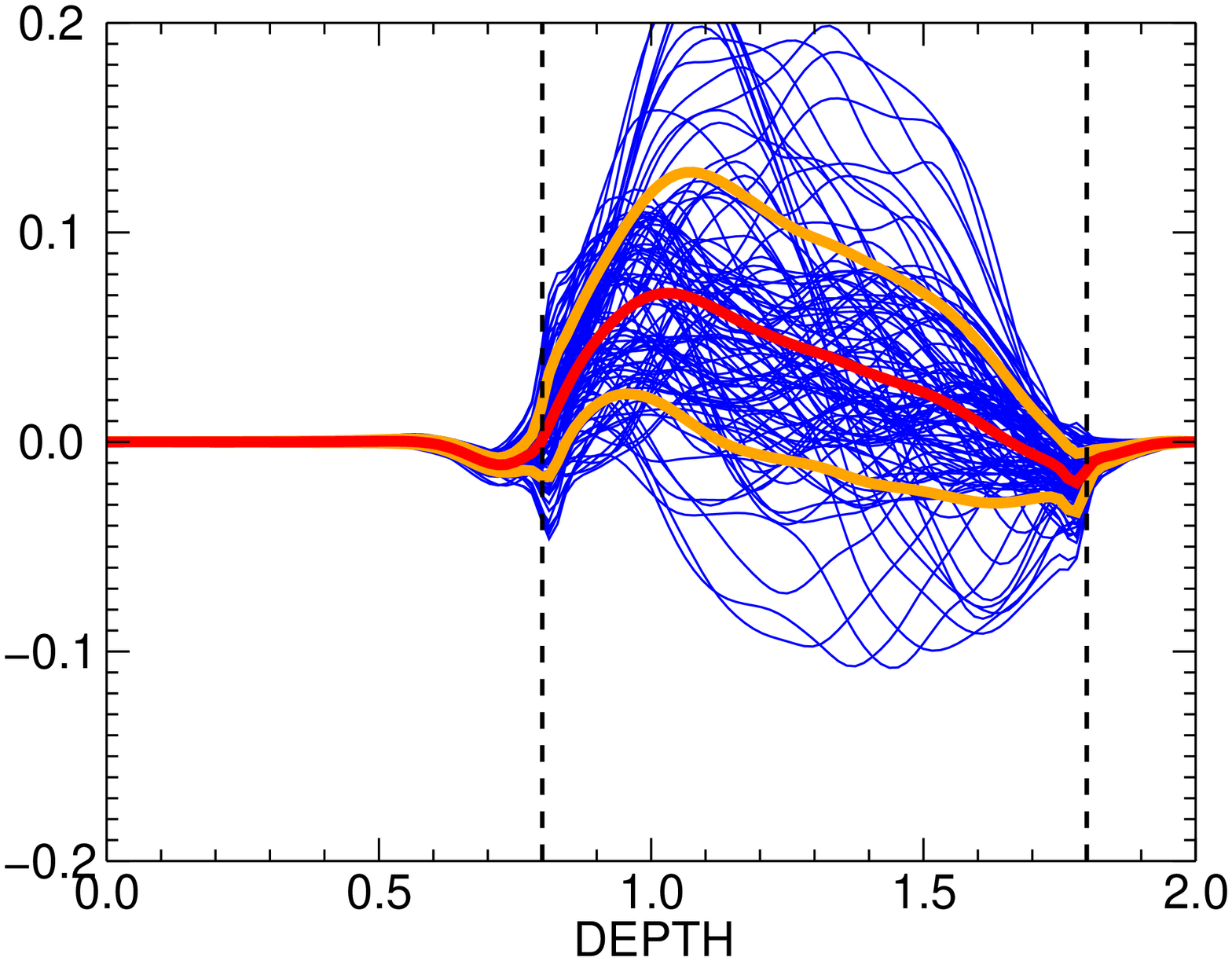}
    \includegraphics[width=0.34\textwidth]{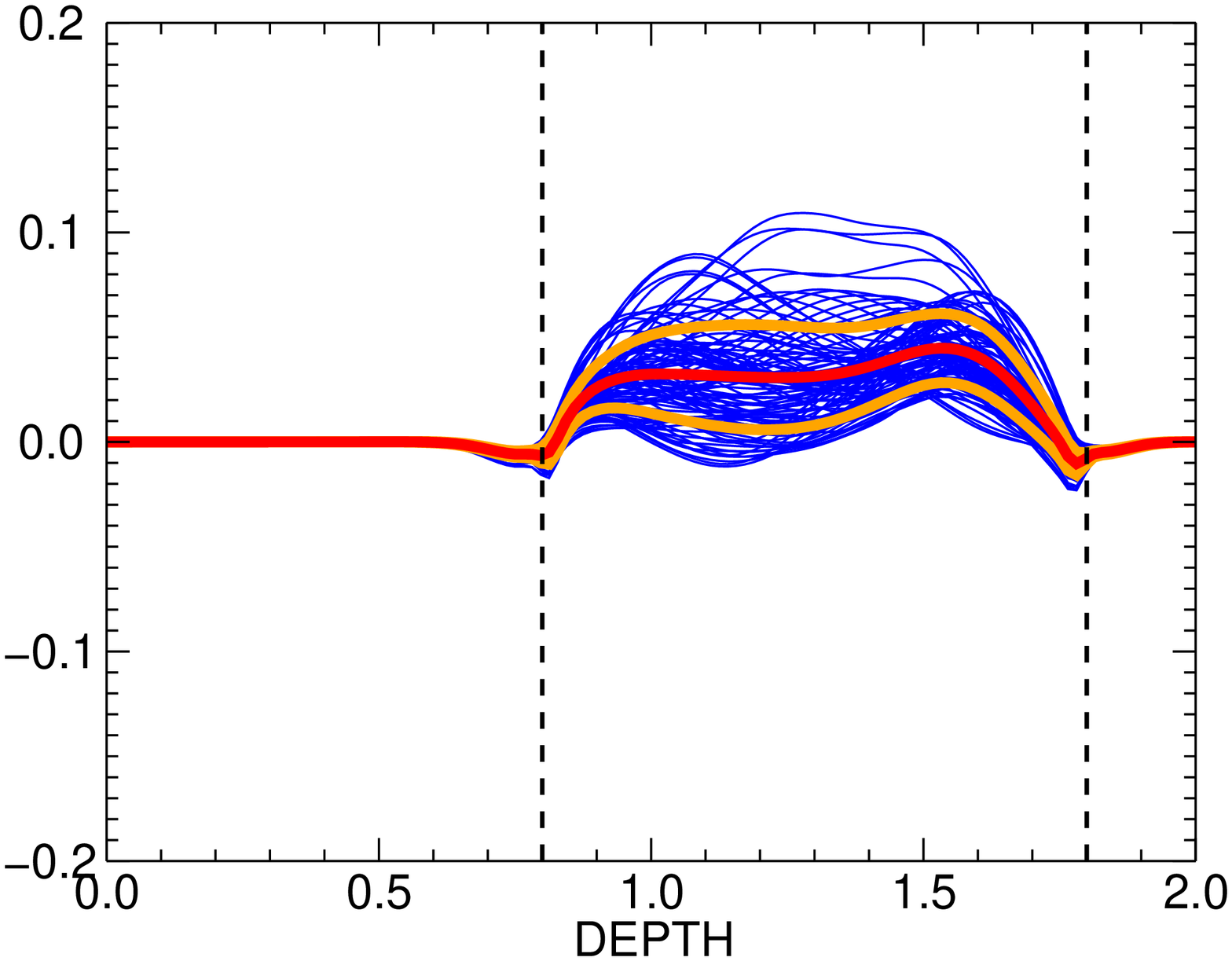}
  }
  \caption{The values of $\kappa$ after  (\ref{corrnew}) for rotating magnetoconvection  with azimuthal magnetic field  $B_\phi=\pm 1$ (left) , $B_\phi= 2$ (middle) and $B_\phi= 3$ (right). $\Om=3$,  $\Pm=0.1$, $\theta=45^\circ$. }
\label{fig2}
\end{figure}

{ { We calculate $\kappa$ } for different magnetic
  background fields for a fixed rotation rate. The right panel of
  Fig. \ref{fig3} confirms that the correlation (\ref{corrnew}) vanishes for
  $\Om=0$. The three examples given in Fig. \ref{fig2} have been computed
  with the rotation rate $\Om=3$ which { corresponds} to a Coriolis number $2
  \tau_{\rm corr}\Om=3.6$.}  The
%
preferred direction $\vec g$ has been fixed to $\theta=45^\circ$
corresponding to mid-latitudes in spherical geometry. { This is a
  natural choice as equator and poles as the two extremes are
  excluded. At the equator we do not expect finite correlations to
  exist while the simulations often meet complications at the poles.
}
%
As expected, the resulting  $\kappa$ is positive and does
not depend on the sign of the magnetic field. It is   about $\lsim
0.1$ for weak magnetic fields. Due to magnetic suppression an increase
of the field by a factor of three reduces the $\kappa$ by the same
factor. An estimation of the analytical result (\ref{kappa}) yields
$\kappa\simeq (1/15) {\rm St}^2$ for $\nu=\meta$. { Hence,
  $\kappa\lsim 0.1$ for Strouhal number unity derived from the analytical 
  expressions is confirmed by numerical calculations.} For the
effective pumping $\hat\gamma=\gamma/\alpha\Om$ the simulations
provide the numerical value of { $\cal{O}$(10)} in (rough) accordance with the
estimates (\ref{alpha2}).

For our argumentation only   standard  mean-field electrodynamics in turbulent media is needed. We note that the   $\alpha$ term in Eq. (\ref{J2}) is turbulence-originated but it  does not need a prescribed  helicity in stratified turbulent media; the helicity parameter $H$ from Eq. (\ref{qu}) does not occur in the calculations. In order to ensure the $\alpha$ tensor being a pseudo-tensor the new  $\alpha$ effect only exists in rotating media which, however, are no longer required to be stratified in density and/or turbulent intensity.

The dynamo number $C_\alpha= |\alpha| R/\etaT$ for large $\Rm$ is
\beg
C_\alpha\lsim\frac{\rm St}{\Rm}\frac{\eta_{\rm rms}}{\meta} \frac{\Om R}{\urms},
\label{C}
\ende
with $R$ the characteristic size of the dynamo domain. That 
{ $C_\alpha$ exceeds unity, which is necessary for dynamo excitation in
$\alpha^2$ models, cannot be excluded for sufficiently rapidly}
rotating large volumes. 
%
 Applying the characteristic values of the geodynamo with $\Rm\simeq 100$ and $\urms\simeq 0.05$~cm/s would  provide $\etarms/\meta\gsim 10^{-4}$ as the excitation  condition
 of an $\alpha^2$-dynamo.  We shall see below that in the { outer
   core of the Earth} such (large) values are not realistic. In the
 solar convection
zone the equatorial velocity $\Om R$  slightly exceeds the maximal
convection velocity but the very large $\Rm$ will prevent sufficiently
high values of $C_\alpha$. The smallness of the presented $\alpha$ effect does not prevent the 
operation of $\alpha\Om$ dynamo models if sufficiently strong differential rotation exists. The standard solutions of these models, however, are oscillating with time scales of the order of the diffusion time.

Finally it might be underlined that (\ref{corrnew}) describes a
general turbulence-induced radial transport of azimuthal
electric current fluctuations which vanishes for $\Om=0$. It exists
for {\em all rotating homogeneous turbulence fields without another
  preferred direction beyond rotation axis and magnetic field
  direction}.
\section{Results and discussion}
We have shown analytically and with numerical simulations that
the eddy-diffusivity in a turbulent fluid is reduced if
not only the flow speed { but also the electric conductivity
  fluctuates}. In this case the effective eddy diffusivity is
smaller than {that without diffusivity fluctuations}. This is
understandable as the large-scale
electric current prefers the high-conductivity islands if they randomly
exist in the fluid.
For small magnetic Reynolds number $\Rm$ the
large-scale diffusivity  decreases  with growing $\eta_{\rm rms}/\meta$
but this effect disappears   for large $\Rm$.
 Figure  \ref{fig1} demonstrates   the reduction effect
as  a phenomenon of (say) a few ten per cent.

If the  fluid  becomes anisotropic in the sense that  one of the components of the flow vector is correlated (or anticorrelated) with the local values of the fluctuating electric conductivity}  then further phenomena appear. Convection  may serve as an example  where the downward and upward flows always  have different temperatures and, therefore, different electric conductivities. If we define in positive radial direction   the correlations as positive then a downward topological pumping of the magnetic field appears in the negative radial direction. With other words, if by the existence of correlations 
the diffusivity fluctuations are   transported in one direction then the magnetic background field is transported in the opposite direction.
This is despite  the fact that the considered turbulence is
homogeneous. Applying the diffusivity  relation $\eta\propto T^{-3/2}$ \citep{S62} to
convection then the correlation $\langle\eta' u'_r\rangle $ (with $r$ as the radial direction in spheres) is negative. It formally describes a downward transport of diffusivity  and hence  the magnetic pumping should go upwards.
The amplitude of the pumping { velocity}, however, is only a few percent of
the turbulent { velocity} which may be smaller by one order of magnitude than
the diamagnetic effect of inhomogeneous turbulence.

The Spitzer relation yields $\eta'/\meta\simeq - 1.5 T'/T$ where a
simple estimate provides $T_{\rm rms}/T\simeq \urms^2/g \ell$ with $g$
as the { acceleration due to gravity}.
The numerical  results $\urms\simeq 0.7$ and $\ell\simeq 0.5$  provide 
$T_{\rm rms}/T\simeq 5\times 10^{-4}$ in excellent agreement with the
outcome of the numerical simulations (Fig. \ref{fig4}). The
fluctuating diffusivity $\eta_{\rm rms}/\meta$ is of the same order
which after (\ref{C}) is consistent with $C_\alpha=\cal{O}$(1) if
{ characteristics of the Earth's core} are applied. However, the very slow
convection flows in the { outer core of the Earth provide} much smaller
values of $\eta_{\rm rms}/\meta\simeq \urms^2/g \ell \lsim
10^{-12}$. 

If the values of our { local convection simulations} are used to compute
(\ref{C}) then $C_\alpha\simeq10^{-5} $, which is far from the
possibility to excite an $\alpha^2$ dynamo.

Also in the bulk of the solar convection zone the temperature fluctuations with $T_{\rm rms}/T\simeq  10^{-6}$ are small. Only  the granulation pattern near   the solar surface exhibits  higher values of order 0.01 \citep{S89}.
\begin{figure}
   \centerline{
   \includegraphics[width=0.45\textwidth]{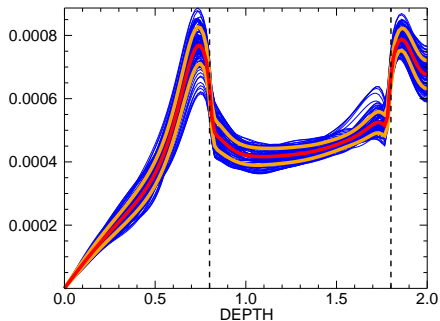}}
  \caption{Similar to the left panel of Fig. \ref{fig2} but for the normalized temperature fluctuation $T_{\rm rms}/T$.  $B_\phi=1$. $\Om=3$,  $\Pm=0.1$, $\theta=45^\circ$. }
\label{fig4}
\end{figure}

All { previously} known turbulence models which produce $\alpha$ effect under the
influence of rotation are inhomogeneous. The turbulence { considered} in the present
paper is homogeneous but anisotropic if molecular diffusivity
fluctuations are correlated (or anticorrelated) with the radial (better: upward and downward) velocity fluctuations. With the vector of the
preferred direction (say $\vec g$) in the turbulence field one can
form a pseudo-scalar $(\vec g\cdot \vec \Om)$ which is needed for
the existence of the pseudo-scalar $\alpha$ in (\ref{alpha}). It is thus challenging to
probe our model for generation of an $\alpha$ effect under the
influence of rotation. Equations (\ref{J2}) and (\ref{intalpha})
represent the analytical results of a quasi-linear approximation. The
coefficient $\alpha$ has the same sign as the diffusivity-velocity
correlation for $\delta$-like frequency spectra but it vanishes for white-noise 
spectra. The ratio $\hat \gamma$ of the pumping term and the
$\alpha$ effect depends on the rotation rate. { We estimate this
  ratio to be $\lsim10$} for Coriolis number unity.

We have probed the properties of the diffusivity-current correlation vector $\langle\eta' \curl \vec{B}'\rangle$ by means of the proxy $\langle u'_r \curl \vec{B}'\rangle$ where it is assumed that the flow component $u'_r$ is correlated (or anticorrelated) with the diffusivity fluctuation $\eta'$. To this end the turbulent flux of electric current (\ref{alpha5}) has been calculated under the influence of rotation and  an uniform magnetic field in azimuthal direction. 
With numerical simulations of { rotating magnetoconvection} driven by very weak
density and temperature stratification the analytical results have
been verified.  The interesting correlations are $\langle u'_r
\curl_\theta \vec{B}'\rangle$ for the pumping term $\gamma$ and
$\langle u'_r \curl_\phi \vec{B}'\rangle$ for the $\alpha$ term. 
We  note that  the  considered turbulence field is non-helical. In opposition to the tensors of helicity  $\langle u'_i \curl_j \vec{u}'\rangle$  and  current helicity $\langle B'_i \curl_j \vec{B}'\rangle$, the current-flux tensor $\langle u'_i \curl_j \vec{B}'\rangle$  is {\em not} a  pseudo-tensor. Both correlations show finite values with the expected
signs and with the correct symmetry properties.
\bibliographystyle{jpp}
\bibliography{superamri}

\end{document}